\documentclass[acmsmall, nonacm=true]{acmart}
\usepackage{rotating} 
\usepackage{array, makecell}
\usepackage{siunitx, mhchem}
\usepackage{tikz}
\usepackage{multirow}
\usepackage{lscape}
\usepackage{longtable}
\usepackage{graphicx}
\usepackage{adjustbox}
\usepackage{pdflscape}
\setcitestyle{square}

\AtBeginDocument{%
  \providecommand\BibTeX{{%
    \normalfont B\kern-0.5em{\scshape i\kern-0.25em b}\kern-0.8em\TeX}}}
\setcopyright{acmcopyright}
\acmJournal{CSUR}

\begin{document}

%
\title[A Holistic View on Resource Management in Serverless Computing Environments]{A Holistic View on Resource Management in Serverless Computing Environments: Taxonomy and Future Directions}

%
\author{Anupama Mampage}
\email{mampage@student.unimelb.edu.au}
\author{Shanika Karunasekera}
\email{karus@unimelb.edu.au}
\author{Rajkumar Buyya}
\email{rbuyya@unimelb.edu.au}
\affiliation{
  \department{Cloud Computing and Distributed Systems (CLOUDS) Laboratory, School of Computing and Information Systems}
  \institution{The University of Melbourne}
  \country{Australia}}

%
\renewcommand{\shortauthors}{A. Mampage et al.}

%
\begin{abstract}
Serverless computing has emerged as an attractive deployment option for cloud applications in recent times. The unique features of this computing model include, rapid auto-scaling, strong isolation, fine-grained billing options and access to a massive service ecosystem which autonomously handles resource management decisions. This model is increasingly being explored for deployments in geographically distributed edge and fog computing networks as well, due to these characteristics. Effective management of computing resources has always gained a lot of attention among researchers. The need to automate the entire process of resource provisioning, allocation, scheduling, monitoring and scaling, has resulted in the need for specialized focus on resource management under the serverless model. In this article, we identify the major aspects covering the broader concept of resource management in serverless environments and propose a taxonomy of elements which influence these aspects, encompassing characteristics of system design, workload attributes and stakeholder expectations. We take a holistic view on serverless environments deployed across edge, fog and cloud computing networks. We also analyse existing works discussing aspects of serverless resource management using this taxonomy. This article further identifies gaps in literature and highlights future research directions for improving capabilities of this computing model.
\end{abstract}

%
%


%

%

%
\maketitle

\section{Introduction}

In computing, resource management refers to the process of provisioning and allocating the right amount of resources for an application, and scheduling the constituent components of the application on the selected resources to meet certain Quality of Service (QoS) goals set by the involved parties. The serverless computing model is built on the principle that the provider takes full responsibility of the entire aspect of resource management for applications, unburdening the consumers of the complexities of infrastructure management. As such, once an application is deployed on a serverless platform, the serverless provider is accountable for the whole chain of activities starting with the initial resource allocation and scheduling responsibilities as well as any subsequent monitoring and resource scaling requirements of the application.

An application deployed on a serverless platform is formed of granular code segments containing the application logic, called "functions", which are generally stateless. A single application may be composed of a single function or multiple functions with dependencies on each other. The user specifies the workflow dependency graph and the abstract resource requirements \cite{jonas2019cloud} and the provider handles the runtime orchestration decisions for the application. A serverless platform could be deployed on distributed computing resources available on multiple infrastructure domains. This could be the public cloud, on-premise private cloud resources or the emerging edge and fog computing architectures. At the edge computing networks, the compute resources will include sensors, smart phones and smart appliances with spare compute and storage capabilities. Fog resources, which form the network between the edge and the cloud, would have compute, storage and memory capabilities at a level in between the edge and cloud layers. The resource management decisions in a serverless platform deployed commercially, need to be directed towards satisfying both the consumer and the provider. QoS goals for the consumer is associated with application performance with regard to the response time, the incurred cost for application execution and concerns of application security. The provider aims to use the underlying resources efficiently, targeting a high throughput system. Increased resource utilization at one point sacrifices the QoS guarantees to the user and thus these are identified as conflicting objectives.

Management of resources in serverless computing environments deployed in cloud, fog and edge resources have been studied by many. The basic scheduling challenge of mapping function execution requests to available compute resources, is influenced by various other challenges that are identified within the broader concept of resource management. These include application workload modelling techniques, mechanisms for application isolation and resource scaling strategies. 

A literature survey can provide a foundation to examine and understand an evolving research area, in the context of existing work. Preliminary surveys on serverless computing identify opportunities of this new computing model, inherent challenges with the serverless concept and ideas for overcoming the said drawbacks \cite{shafiei2019serverless}, \cite{jonas2019cloud} \cite{fox2017status}. Hellerstein et al. \cite{hellerstein2018serverless} pinpoint specific challenges of the serverless model with regard to applicability for data-driven distributed computing. Garcia et al. \cite{garcia2019servermix}  analyse and discuss trade-offs of today's serverless platforms required for effective processing of big data analytics applications. Many studies focus on exploring performance of existing commercial and open-source serverless platforms \cite{wang2018peeking}, \cite{lee2018evaluation}, \cite{kritikos2018review}. Eismann et al. \cite{eismann2020review} present a characterization of serverless use cases. None of these works are focused specifically on the overall aspect of resource management under the serverless computing model and the emerging body of related literature.

In this article, we propose a broad survey on the aspects of resource management in serverless computing environments covering the cloud, fog and edge computing environments. We first identify three major elements of resource management under this computing model, namely application workload modelling, resource scheduling and scaling. We then propose a taxonomy of factors which affect these three aspects, considering the unique characteristics of this novel computing model. This classification covers effects of serverless system designs, application characteristics, and QoS goals, on the application workload modelling, scheduling and scaling decisions, referring to existing serverless platforms as well as research literature over the past few years. We then analyse key techniques for resource management in existing literature by applying this taxonomy. This survey presents serverless system designers with a broader overview of the primary characteristics to be considered for their target infrastructures, a clear idea of the influencing factors and a summary of existing approaches for researchers studying resource management techniques for serverless environments and ways to maximize benefit from their deployments, for serverless application developers.

The rest of the paper is organized as follows: Section 2 provides a brief background on the serverless model, its unique features, an overview on the existing serverless platforms and frameworks, a highlight on the primary use cases for this computing model and a discussion on the concept of serverless resource management which leads to the motivation for this work. Section 3 presents the proposed taxonomy of resource management. Section 4 summarizes existing works on serverless resource management based on the taxonomy. Section 5 concludes the paper by summarizing our detailed review and highlighting ideas for future directions in this area, stemming from the gaps identified in existing approaches.

\section{Background}

The scope of resource management in any computing environment needs to be aligned with the specific challenges unique to that particular ecosystem. In this survey we take a holistic view on the different aspects of the serverless computing  model, that influence designing resource management techniques that enable stakeholders to meet their respective goals and objectives.

In order to better understand the scope and breadth of the resource management problem, in this section, we present a brief background on the concept of serverless computing, the basic execution flow and the key characteristics of a serverless environment. Further, we provide a summary on the features of existing key commercial and open-source serverless platforms, along with a short introduction to the main application domains for serverless use cases. Then we introduce the scope of resource management under this computing model, its challenges and the motivation for our work.

\subsection{Serverless Computing}

Serverless computing is an application service model in which, the provider manages the server, and handles all the responsibilities related to the execution of the application during its lifetime, with minimum involvement of the user.

\begin{figure}[h]
	\centering 
	\includegraphics[width=\linewidth]{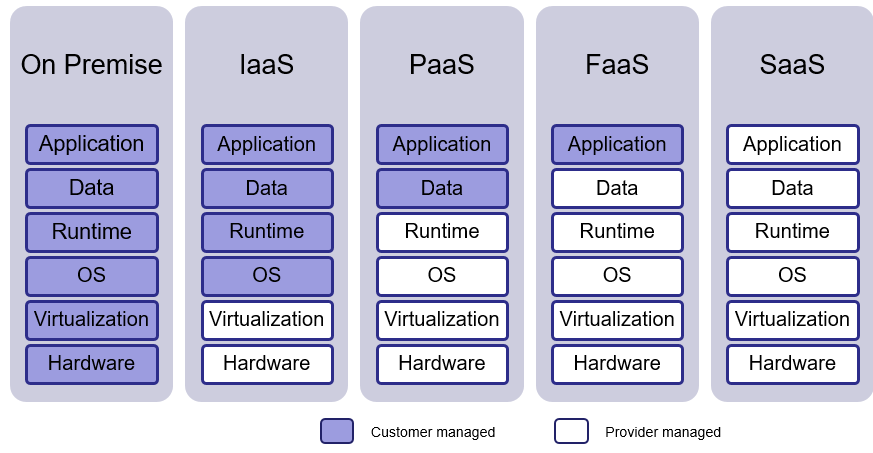}
	\caption{Evolution of Cloud Service Models}
	\label{fig:ServerlessEvolution}
\end{figure}

Starting with bare-metal servers maintained on-premise at redundancy, in order to enable high availability, the way server infrastructure is managed for use by consumers has evolved largely over the years. The biggest transformation in this regard was with the advent of the concept of cloud computing around early 2000. Along with it came the three cloud service models, Infrastructure as a Service (IaaS), Platform as a Service (PaaS) and Software  as  a  Service  (SaaS). Under the IaaS model (e.g.: AWS EC2, Azure VMs, Google Cloud Platform), the provider manages hardware resources at their data centers. The developer rents out required virtualized cloud resources from a vendor and thus possesses the responsibilities of resource provisioning, runtime configurations and the management of application code and data. While  this  pay-as-you-go  model  in  acquiring  cloud  computing  resources allows customers to pay for only the resources leased from the cloud providers, studies show a significant gap between the resources acquired and paid for by cloud users and the actual resource utilization (CPU, memory  etc.) \cite{castro2019rise}. Thus an inherent challenge with the IaaS model is the resultant under utilization of resources in general, when resources are acquired to match peak demand of a system, and the resultant over utilization and performance degradation when resources are acquired to cater to average demand levels. The PaaS model (e.g.: AWS Elastic Beanstalk, Azure app services, Google app engine) provides a platform for users to develop, run and manage customized applications while the execution environment is managed by the cloud provider. The serverless computing model is a similar paradigm where, while the developer has control over the code they deploy, they need to follow certain standards to suit the provider platforms. In addition, this model offers far more granularity with regard to application scaling and billing schemes which differentiate it from other execution models and create new opportunities. For example, under the serverless model, an application has the option to scale to zero, with no instance of the application consuming any resources when there is no traffic. This is in contrast to a PaaS model, where at least one instance of the application will always be up and running, consuming some amount of resources, irrespective of the traffic levels.

\subsubsection{Serverless Computing Architecture}

\begin{figure}[!h]
	\centering 
	\includegraphics[width=\linewidth]{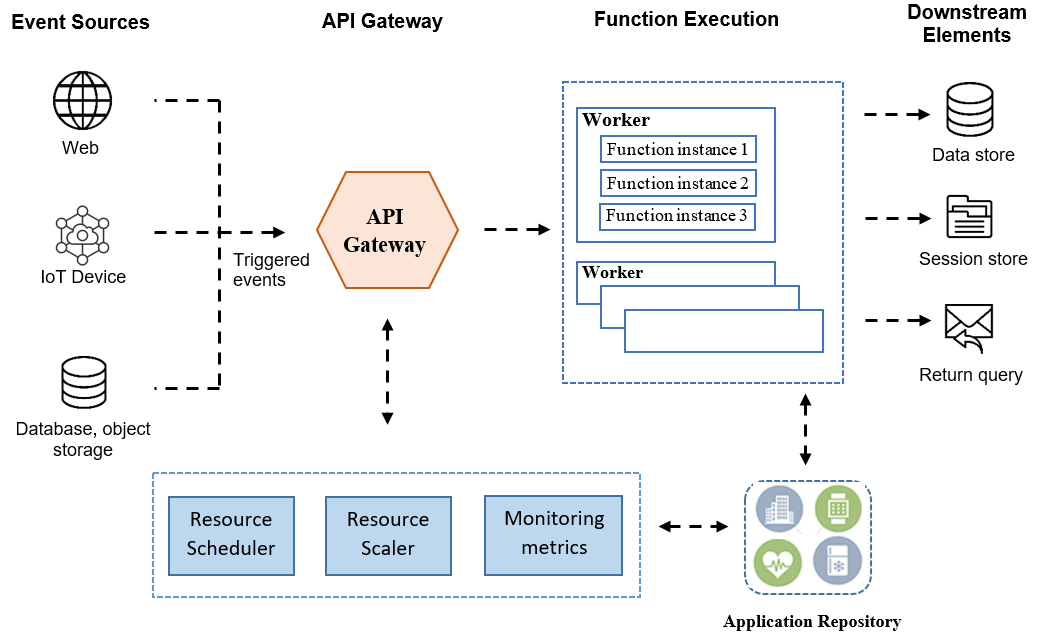}
	\caption{Serverless Architecture}
	\label{fig:ServerlessArchitecture}
\end{figure}

Serverless architecture is an event-driven architecture where users would initially deploy code with the application logic, in the form of stateless functions. A serverless platform defines a set of event sources which could trigger the invocation of these pre-deployed functions as per the user requirement. A user defines rules, binding deployed functions with corresponding event sources. The supported event sources depend on the platform and these could be HTTP requests from a user interface, a change to a database or an object storage, a notification from an Internet of Things (IoT) device etc. Figure \ref{fig:ServerlessArchitecture} illustrates the basic execution flow of the serverless architecture.

Upon the occurrence of an event, a request(s) is sent to the API gateway in order to invoke the relevant function(s) as per the defined set of rules. The scheduler then determines which worker node is best suited to accommodate the function execution and dispatches the execution request to the relevant worker. A suitable isolated environment with the required resource configurations (e.g.: a container) is created on the worker to accommodate the request. Function execution commences once the required runtime and the associated application code from the application repository, are loaded on to the created environment. Upon completion of execution, the response is sent to the user and the environment created is usually destroyed, releasing the allocated resources. Intermediary data and state management is done via external storage services. The function scaling decisions to meet the requirements of subsequent function execution requests, are taken considering the data from monitoring metrics and the implemented function scaling logic.

\subsection{Key Characteristics of Serverless Platforms}

As per \cite{baldini2017serverless}, below are some unique key features of existing serverless platforms.

\begin{itemize}
    \item Auto-scaling – The platform is expected to be able to scale resources automatically and instantly as per the demand. Serverless platforms are equipped with container technologies which have minimal start-up delays, thus enabling the provision of thousands of instances within a few seconds. Similarly, when there is no traffic to an application, the function instances scale to zero maintaining minimum idle resources.
    \item Billing – The usage is metered and users are only charged for the resources used when serverless functions are in execution. This means that when a function scales to zero and no node is running the user’s code, there is no cost to the user.
    \item Performance and limits – A variety of limits are set on the run time resource requirements of a serverless code, including the number of concurrent requests, maximum memory and CPU resources available to a function invocation and an upper bound in execution time before a function instance times out.
    \item Programming languages – Most platforms support function code written in multiple programming languages which include Javascript, Java, Python, Go and Swift.
    \item Security and accounting – Serverless platforms are multi-tenant and thus providers need to be considerate of isolating the execution of functions of different users. Linux kernel features like namespaces and cgroups offered by container technologies provide some level of resource isolation for individual function executions.
    
\end{itemize}

\subsection{Serverless Computing Platforms and Frameworks}

\begin{figure}[h]
	\centering 
	\includegraphics{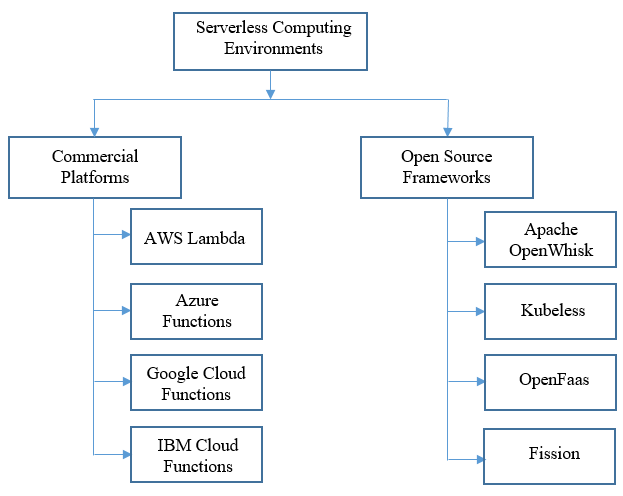}
	\caption{Existing Serverless Platforms and Frameworks}
	\label{fig:ServerlessPlatforms}
\end{figure}

Currently, all the major cloud service providers have launched commercial serverless platforms, namely Amazon Web Services (AWS) Lambda, Google Cloud Functions, Azure Functions and IBM Cloud Functions. While they provide additional complementary services as well for serverless application executions, they require the function code to be composed in certain ways resulting in vendor lock-in in the long term. To overcome these limitations, open source serverless frameworks have emerged over the years. These frameworks could be deployed on private cloud resources as well as on devices at the edge/fog, thus bringing the serverless computing capabilities on-premise with better flexibility. Figure \ref{fig:ServerlessPlatforms} presents the existing commercial serverless platforms and some of the open source frameworks.

\subsubsection{AWS Lambda}

Function as a service (FaaS) offering from AWS is done via the AWS Lambda serverless platform launched in 2014. AWS lambda currently holds the leading position in the market for serverless computing with a wide range of services available. Lambda supports code written in Node.js, Python, Java, C\# and Golang languages \cite{lee2018evaluation}. Lambda identifies changes to data in an Amazon Simple Storage Service (Amazon S3) bucket or an Amazon DynamoDB table, HTTP requests using Amazon API Gateway and API calls made using AWS SDKs as event triggers. AWS offers a free-tier of 1 million requests and 400,000 GB-Seconds (GB-s) of computing time per month. Beyond this limit, function executions are charged per GB-s of memory allocated and the number of execution requests. Lambda allows the developer to specify a maximum memory limit that the function will have access to, at the deployment stage, and allocates CPU power proportionate to the allowed memory limit \cite{AWSLambd95:online}. AWS enables applications to be executed as composed of a single function or multiple functions. AWS Step Functions \cite{AWSLambd95:online} facilitates developers to define workflows with a sequence of Lambda functions and create a state machine that orchestrates the set of functions in the application. AWS uses a new virtualization technology called Firecracker, to execute function requests. These lightweight micro-virtual machines (microVMs) have a similar level of workload isolation  and security provided by traditional VMs and the resource efficiency of containers \cite{Firecrac95:online}. AWS Lambda treats instance placmeent in VMs as a bin packing problem to maximize VM memory utilization. AWS has set a concurrency limit for scaling a single function and experiments show that idle instance recycling is done based on a pre-defined inactive period \cite{wang2018peeking}.

\subsubsection{Azure Functions}

Microsoft launched its serverless services in 2016 as Azure functions. Azure supports a number of runtime languages such as C\#, Node.js, PHP, Bash, Power Shell. The billing scheme is similar to that of AWS except that billing is done per GB second of average memory consumed during an execution instead of the memory allocated. Azure uses the concept of function app as the unit of deployment and management of a serverless application. A function app is comprised of one or more functions which are deployed, managed and scaled together. All the functions in a function app share the same pricing plan and runtime configurations  \cite{AzureFun89:online}. Azure Functions seems to try not to co-locate concurrent instances of the same function on the same VM, which indicates a spread placement approach \cite{wang2018peeking}. Azure currently supports the execution of stateful functions as well in a serverless computing environment with Azure Durable Functions, which is an extension of Azure Functions \cite{AzureFun89:online}.

\subsubsection{Google Cloud Functions (GCF)}

Google released its serverless solution in 2017. GCF supports code written in Node.js only at the moment. Google has a free tier of 2 million requests with 400,000 GB-Seconds (GB-s) of computing time per month. Their pricing scheme is different to AWS Lambda and Azure Functions pricing mechanism since the user is charged for both GB-s of memory provisioned and GHz-s of CPU provisioned \cite{CloudFun60:online}. GCF supports a set of primary triggers and additional triggers and the user application is allowed to be integrated with any Google service, supporting cloud Pub/Sub or HTTP callbacks.

\subsubsection{Apache OpenWhisk}

OpenWhisk is an open source serverless platform developed by IBM and later incorporated as an Apache incubator project. It represents the underlying technology used in IBM Cloud Functions. OpenWhisk combines technologies such as Nginx, Kafka, Docker and CouchDB in forming its serverless platform. OpenWhisk supports many deployment options and could be deployed both locally and within a cloud environment. The OpenWhisk programming model is based on the three primary concepts, actions, triggers and rules. Actions are functions that execute deployed code. Triggers are a set of events created from different sources. Rules bind actions with triggers. OpenWhisk supports a number of runtimes including, .Net, Go, Java, JavaScript, PHP and Python \cite{Document62:online}. In selecting a VM instance for a function execution, OpenWhisk follows a hash-based first-fit heuristic where the function name's hash value is used to identify a preferred host in order to improve reusing warm containers \cite{stein2019adaptive}. Multiple functions, which may even be implemented in different languages, could be composed together to create a function pipeline called a sequence. A sequence could be considered a single action in terms of it's creation and invocation. In executing a function sequence, the output from one action becomes the input to the next action in the sequence \cite{Document62:online}. OpenWhisk makes use of pre-warmed containers and warm containers (container re-use) to manage unwanted container startup delays.

\subsubsection{Kubeless}

Kubeless is a Kubernetes-native open-source serverless framework developed by Bitnami. Kubeless creates functions as a custom Kubernetes resource using a Custom Resource Definition (CRD). An in-cluster controller is used to watch these custom resources and execute functions on-demand by dynamically launching runtimes as required. Kubeless supports runtimes of Golang, Python, NodeJS, Ruby, PHP, .NET and Ballerina and allows to use HTTP or event triggers to invoke functions. The Kubeless framwork uses core Kubernetes functionalities such as deployments, ConfigMaps and services as it is. It also leverages the native Kubernetes components for function scheduling, auto-scaling and monitoring as well.

\subsubsection{OpenFaas}

OpenFaas started as an independent project by Alex Ellis in 2016. Initially, it was developed in collaboration with VMWare and now it involves a large community of users and developers. OpenFaas framework is built on docker and Kubernetes and could be deployed in a private or public cloud environment, and even on a resource constrained edge device such as a Raspberry Pi, due its lightweight nature. OpenFaas provides an API gateway for invoking functions, which can be accessed via its REST API, Command Line Interface (CLI) or the User Interface (UI). The gateway acts as an external route to the functions, collects cloud native metrics through Prometheus, and also takes function scaling decisions with the help of Prometheus and an AlertManager component. AlertManager works by reading the requests per second metric from Prometheus and alerting the gateway to scale functions based on the min/max replica count set at function deployment. Alternatively users could use the built-in Horizontal Pod Autoscaler (HPA) of Kubernetes. OpenFaas offers runtimes of Node.js, Python, and Go for function deployment. OpenFaas also supports the orchestration of multi-function workflow applications with synchronous and asynchronous function chains and parallel branching.

\subsection{Application Domains}

In general, bursty and compute-intensive workloads which are stateless and ephemeral could benefit more from a serverless architecture. From a cost perspective, the auto-scaling feature of the platforms proves useful, when traffic arrives in bursts (inconsistent traffic levels), since the system can also scale to zero when there is no traffic \cite{baldini2017serverless}. During recent times, serverless computing has increasingly being explored for use in many applications domains such as web and mobile, big data, internet of things, machine learning model training and large scale mathematical computations. The nature of the serverless workload will vary depending on the different requirements of these domains and their inherent characteristics. The focus of the resource management techniques need to be adapted accordingly. As discussed in the following sections, the core design features of the existing platforms make the serverless model easily adaptable for some domains while in others, an extra effort is needed to reap the full benefits of this novel model.

\subsubsection{Web Services}

Serverless model is said to have been initially developed for lightweight use cases like serving APIs or small backend services \cite{kuhlenkamp2019evaluation}. Even today, studies on serverless use cases reveal that webservices is the most common application domain utilizing this new computing paradigm due to ease of adoption \cite{eismann2020review}.

\subsubsection{Big Data Analytics}

Due to high data volumes and computational requirements, big data processing is traditionally undertaken using clusters of machines with jobs consisting of multiple tasks being executed across a set of machines. Advancement of technology in the field has addressed these performance and scalability needs through distributed computing frameworks specialized for big data analytics \cite{spark2018apache}, \cite{carbone2015apache}. The costs and the knowledge required for configuring, deploying and maintaining these systems is still a challenge. The lower startup times, automated resource management, function level auto-scaling and granular billing features present an interesting opportunity in the serverless model for big data processing \cite{kuhlenkamp2019evaluation}. The potential of serverless for big data analytics is being recognized by the major cloud providers as well. AWS provides guidance on reaping benefits of Lambda for streaming data analytics \cite{AWSLambd95:online}. IBM introduces IBM-PyWren, a data analytics platform using IBM Cloud Functions \cite{sampe2018serverless}. A number of research efforts are seen trying to adopt the serverless model for obtaining favorable results for big data applications. A prototype serverless spark execution engine using AWS Lambda as the spark cluster is presented \cite{kim2018serverless}. FaaS model is explored for various MapReduce jobs \cite{gimenez2019framework}, \cite{enes2020real}. However, many fundamental challenges still exist that impede the performance of distributed data-driven applications on current serverless platforms. The data-shipping architecture where function execution is completely isolated from data, and the short-lived, stateless nature of functions makes it a requirement to routinely ship data to code, for each function invocation. Functions are also usually not network-addressable and thus two functions could only communicate through slow and expensive storage. Current FaaS platforms lack access to any specialized hardware and functions are only provisioned with the processing power of a timeslice of a CPU hyperthread \cite{hellerstein2018serverless}.

\subsubsection{Internet of Things}

FaaS model could be beneficial for adoption in IoT applications in edge/fog computing networks owing to a number of factors. Deploying applications as a number of lightweight functions go in line with resource limitations at the edge devices. Statelessness of serverless functions add portability for parts of applications to be moved across the edge/cloud computing network with lesser complications. As a result, today, serverless computing has been exploited in many IoT domains including home automation and other custom-built IoT solutions. AWS offers AWS IoT Greengrass \cite{AWSIoTGr47:online} which is included as a service in the serverless ecosystem as well and this allows processing data at the edge. Azure IoT edge \cite{AzureIoT52:online}, which could be used in conjunction with their FaaS service Azure Functions, moves cloud analytics to the edge devices. A number of researches focus on introducing frameworks and scheduling techniques for serverless applications deployed in edge-cloud computing networks \cite{elgamal2018costless}, \cite{bermbach2020towards}, \cite{pinto2018dynamic}, \cite{das2020performance}, \cite{baresi2019towards}, \cite{cheng2019fog}. To date, challenges do exist in adopting the serverless model in IoT scenarios. Retrieval of large data sets to the edge dynamically, each time a function is called is not practical specially with latency sensitive applications. Thus arises the need for a caching service at the edge. Geographic dispersion, heterogeneity and affinity of data sources at the edge suggest the need for more decentralized resource pooling and scheduling techniques devised specially for IoT application scenarios.

\subsubsection{Machine Learning}

Machine learning (ML) applications typically consist of three phases: model design, model training and model inference (model serving). VM clusters were traditionally used for the diverse tasks in ML model training. The distinct stages of a ML workflow pipeline consist of varying computational requirements. As such the traditional approaches face several challenges such as the need for developers to provision, configure and manage these resources and the over and under provisioning of VM resources during different stages of the execution of a training workflow \cite{jonas2019cloud}. On the face, serverless computing seems to be a promising approach for resource provisioning challenges for ML users, in terms of its simplified deployment opportunities, fine-grained resource provisioning and billing models and the ability to auto-scale both computation and storage resources. However, the small local memory and storage footprints of serverless functions, lack of function to function communication or fast shared storage, the short-lived function runtimes and unpredictable launch times make serverless platforms less conducive for ML processing as it is \cite{carreira2018case}. Research is being undertaken in this area to propose serverless frameworks with potential developments that could overcome these specific limitations. Carreira et al. \cite{carreira2019cirrus} introduce a framework for ML training pipelines on serverless computing. Results outperform a VM-based execution in terms of training completion time, but at a relatively  higher cost. Out of the three phases of ML applications, Ali et al. \cite{ali2020batch} identify model serving as having the greatest potential to benefit from serverless, since prediction request arrival is usually dynamic and the requests have strict latency requirements. They show that by devising a strategy to enable batching (i.e., bundling of several ML inference requests and serving them together), ML serving inference process could benefit significantly from serverless.

\subsubsection{Mathematical Computation}

Large scale mathematical computations are traditionally deployed on supercomputers or high- performance computing clusters connected by high-speed, low-latency networks \cite{jonas2019cloud}. Considering this, serverless seems a poor fit for such applications. However, the ability to unburden non-computer scientists of having to manage infrastructure and scalability to support varying needs of resource parallelism during a computation, highlights benefits of a FaaS model over managing a cluster with a fixed size. Shankar et al. \cite{shankar2018numpywren} present Numpywren, a serverless system for linear algebra computations. Werner et al. \cite{werner2018serverless} present a prototype for matrix multiplication using FaaS. These experiments show that serverless computing could be a good fit for large scale linear algebra (e.g.: matrix multiplication, singular value decomposition) when computation time dominates communication delays. But the high latency of external storage causes limitations for smaller problem sizes. 

\subsection{Resource Management in Serverless Computing Environments}

The core differentiator of the serverless model from other computing models is the complete shift of server management responsibilities to the vendor, thus rendering the model ’serverless’, from the perspective of the developer. A major, if not the primary aspect of server management, lies in the process of proper management of server resources for the execution of applications. Resource management in a serverless environment refers to the overall aspect of managing the resource requirements of an application workload and the available system resources efficiently, with minimal involvement of the user. Due to the autonomic nature of the expected resource management process in these environments, special focus is required on each step of this process for better performance of the applications and the system. We identify three major aspects of resource management, which need to be dealt with in a manner suitable for this new serverless computing model.

\begin{itemize}

\item {Application Workload Modelling}: Developers prefer minimal work in using  a serverless deployment model. Hence, instead of having to specify a resource configuration and other application characteristics when deploying an application, it is ideal for a serverless platform to be able to infer application characteristics using workload modelling and prediction techniques. An effective scheme for understanding characteristics of an application, enables developing better resource scheduling and scaling techniques as well, which help in satisfying user QoS requirements.

\item {Resource Scheduling}: Mapping workloads to suitable host nodes based on their resource requirements, while efficiently utilizing the available resources is an important challenge that is of interest to both the developer and the cloud providers or system owners. Scheduling also involves determining the order of  execution of applications when the resource demand exceeds the available resource capacity. While the developer would expect certain QoS guarantees, it is essential for the provider to manage resources efficiently, which is a primary goal of resource management.

\item {Resource Scaling}: Under the serverless model, environment creation and resource allocations to applications happen in real time, as and when workloads arrive. This ensures greater flexibility in resource allocations and higher resource efficiency. In order to maintain required application performance while maintaining scaling at such a granular level requires smart and dynamic resource scaling techniques. 

\end{itemize}

\subsubsection{Challenges in Resource Management}

Below we have identified challenges associated with resource management strategies in a serverless environment.

\begin{itemize}
    \item Cold start delay: Due to the auto-scaling nature of these environments, resource creation needs to happen on the go and setting up new resources for a function execution results in a considerable start up time. Applications often face performance degradations due to this initial delay, which becomes specially significant for functions with very short execution times.
    
    \item Co-located application interference: Applications deployed on serverless platforms run in multi-tenant environments, inside specially created isolated environments, such as containers. When several applications run on the same host node and compete for the same set of resources, it is difficult to fully avoid resource interference effects on the applications. Developing sandboxing techniques which are light-weight enough to avoid large set up times and secure enough to provide the required level of resource isolation is an associated challenge.
    
    \item Resource Efficiency: In contrast to a general cloud computing billing model, under the granular serverless billing model, users are charged for only the resources allocated or consumed during the application execution, with a millisecond accuracy. The provider on the other hand may be maintaining the underlying infrastructure for longer periods. This creates the need for special attention to have strategies for high resource efficiency on the host nodes.
    
    \item Diverse Workload Management: Due to minimal user involvement in the server resource management process, serverless systems need to develop an understanding on the application and workload characteristics on their own, in order to deliver a favorable outcome.  The diverse nature of applications that are being deployed on serverless platforms makes this a challenging task. On the other hand, lack of an understanding on the application resource requirements, application model and workload arrival patterns could results in delays in resource set up, heightened resource interference effects, etc. that lead to customer dissatisfaction. 
    
    \item Runtime limitations: To alleviate the risk of loosing flexibility of the infrastructure, serverless providers impose various runtime limitations on applications. These include limitations on the maximum allocated CPU, memory, disk, I/O resources as well as maximum allowed execution time for a function instance. Platforms also configure an upper limit on function concurrency levels which limits the number of parallel executions of the same function by a user. Although successful in preserving flexibility of the system, these limitations deem the serverless platforms unsuitable for certain applications.

\end{itemize}

\subsubsection{Motivation}

The three aspects of resource management identified above need to be addressed considering the aforementioned associated challenges. Various techniques have been experimented by researchers to overcome these challenges and devise better workload modelling, resource scheduling and scaling techniques. In this paper, we identify a classification of the factors which influence the successful management of resources in these environments by reviewing the existing techniques for resource management. In forming this classification, we take into account the unique challenges of this model identified above, in addition to discussing the general concerns of resource management in the context of the serverless model.

Successfully undertaking resource management in any system is determined by the underlying system design features as well as the understanding on the characteristics of the incoming workloads. As such, our classification comprises of a reference to the key concerns of designing a serverless system with regard to the aspect of resource management,  characteristics of the workloads submitted to these systems and how they affect resource management decisions, and finally, the primary goals of resource management in these environments. Further, we summarize the existing research work related to workload modelling, resource scheduling and scaling techniques in the serverless domain, using our proposed taxonomy. We also propose ideas for future work in developing enhanced techniques. Serverless system designers would benefit from our classification by gaining an understanding on the key focus areas of a system design under this computing model and also the existing approaches that have already been evaluated. Researchers studying resource management techniques would be able to refer to existing approaches which could subsequently form the basis for the design of novel and rigorous techniques in the future. This classification would also assist application developers in the design of their serverless applications, choosing the right infrastructure and in budgeting their deployments. The main contributions of this work are as follows:

\begin{itemize}
    \item We identify three major elements of resource management in serverless environments
    \item We propose a taxonomy of the factors which influence resource management under this computing model, by reviewing existing literature on the above identified elements.
    \item We summarize the existing works on serverless resource management in the context of the proposed taxonomy.
    \item We identify research gaps in different aspects of serverless resource management and propose ideas for future research directions.
\end{itemize}

\section{The Taxonomy}

\begin{sidewaysfigure}
\begin{center}
    \includegraphics[width=\textwidth, trim = 0 15cm 0 0]{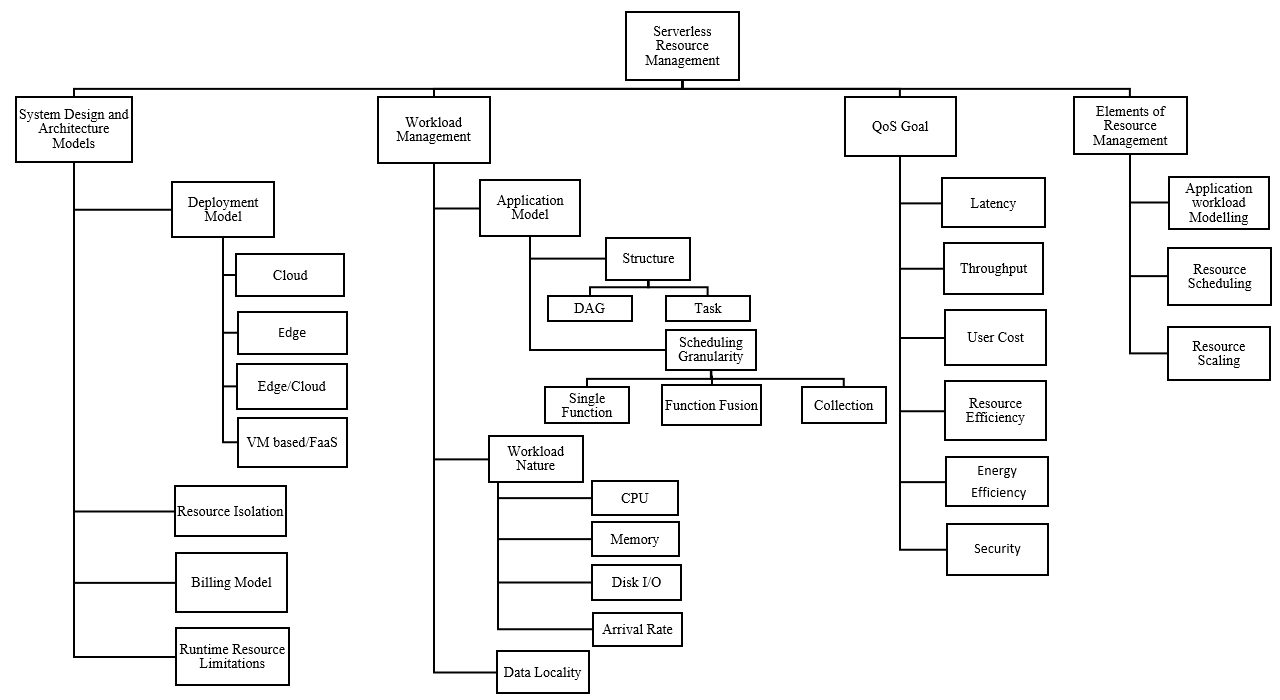}
	\caption{The taxonomy of resource management in serverless computing environments}
	\label{fig:Taxonomy}
\end{center}
\end{sidewaysfigure}

The top level of our taxonomy is comprised of: systems design and architecture models, workload management, Quality of Service (QoS) goal and the key elements of resource management. Figure \ref{fig:Taxonomy} illustrates the proposed taxonomy. We discuss each category in detail in the following sections.

\subsection{System Design and Architecture Models}

The design of a serverless platform needs to address the inherent unique features offered by this computing model and also the associated key challenges. We characterize these factors in terms of the platform deployment model, resource isolation techniques, the incorporated pricing models and the nature of imposed runtime resource limitations. Identifying these influencing factors help in developing effective algorithms, system models and other techniques for the efficient management of resources in serverless environments.

\subsubsection{Deployment Model}

Serverless platforms could be deployed on public cloud, private cloud or on edge resources. The serverless model could also be used in conjunction with a serverful model, i.e.: a VM based deployment. The deployment model is concerned with the nature of infrastructure combination from each resource environment, the associated resource capabilities, pricing models and subsequent resource management decisions suited for each environment for better QoS. This section is focused on cloud and edge only deployments as well as hybrid deployment models with resources used interchangeably from different environments for better performance and efficiency.

\par $\bullet$ \textit{Cloud}: This is the most common as well as the basic deployment model targeted by serverless computing when it first emerged. A cloud environment could be an on premise, user managed, private environment or a vendor managed public environment with seemingly unlimited resources. A private cloud environment gives better flexibility to the user to device scheduling techniques to achieve desired performance guarantees. In addition, privacy is less of a concern. While these environments usually enable reaping cost benefits in the long run for regular workload scenarios, the downside is the inability to meet sudden demand surges if additional hardware is not maintained. With a private/public hybrid cloud model, whenever the load goes beyond the capacity of local infrastructure, some of the functions could be dispatched and processed in the public cloud at a cost. In a hybrid model, one could use a commercial serverless provider's service together with an open source serverless framework deployed on a private network. The serverless scheduling challenge then is to decide when and which functions are to be off-loaded to the public cloud for better QoS guarantees and cost benefits. These decisions need to consider the data transfer times across networks, resource set up times, along with the load levels and required QoS guarantees. Das et al. \cite{das2020skedulix} propose a hybrid cloud scheduling framework for multi-function serverless applications with AWS Lambda as the public cloud and OpenFaaS \cite{Introduc35:online} deployed in the private cloud.

\par $\bullet$ \textit{Edge}: Edge computing leverages computing power and storage facilities close to the consumer in order to reduce application latencies and bandwidth usage. As discussed previously under serverless use cases, serverless computing seems to be a good fit for the deployment of applications across resource constrained edge computing networks due to the ephemeral nature and portability of serverless functions. Gand et al. \cite{gand2020serverless} propose an architecture for clustered container applications for the edge, based on serverless computing. Baresi et al. \cite{baresi2017empowering} propose a serverless edge computing architecture for mobile applications with low latency and high throughput requirements. They use mobile devices and mobile edge computing servers as their main computational elements and evaluate the feasibility of their approach via a mobile augmented reality use case. Containers which are often used as the isolation mechanism for serverless functions, are at times not viable for edge environments with limited resource capabilities due to their inherent set up and runtime overheads. Hall et al. \cite{hall2019execution} propose a novel run time and an isolation mechanism for serverless functions called WebAssembly which reduces resource startup times and the resource provisioning requirements. 

\par $\bullet$ \textit{Edge/Cloud}: A hybrid deployment model leveraging resource capabilities of both the edge and the cloud is often times the most viable solution under many practical scenarios. But many challenges exist, for realizing the true benefit of such a setting. For an application deployed in a serverless edge/cloud infrastructure, which functions are to be deployed on the fog or cloud resources and which node containing the deployed function is best suited for accommodating a new request is a scheduling decision. The resource allocation and scheduling decisions for serverless applications in an edge/cloud computing network will depend on how compute intensive the function is, the size of data involved and data transfer costs over different network paths, the cost and the setup time of resources at each location, and the expected QoS levels \cite{elgamal2018costless}, \cite{pinto2018dynamic}, \cite{bermbach2020towards}.

\par $\bullet$ \textit{VM based/ FaaS}: Traditionally VM based deployments were used to dynamically scale resources as per the demand. However challenges exist such as higher costs due to over-provisioning or SLA violations due to under-provisioning. Under the serverless model, functions are auto-scaled within containers which have a low startup latency. Further, user is billed per function invocation at a very granular level, thus avoiding resource over-provisioning costs. However, it is observed that deploying an entire application as serverless functions would not be cost effective at times. Lambda functions are expensive (cost increases linearly) when there is a fixed load (constant request arrival rate for a function) and a higher average request rate. Thus it is seen that functions are more cost efficient with demand variations and lower average request rates while VM based deployments server better with higher arrival rates. Exploring hybrid models of VM and FaaS deployments is an interesting dimension \cite{gunasekaran2019spock}.

\subsubsection{Resource Isolation}

Serverless systems are multi-tenant systems and thus techniques for isolating applications from each other is important for reasons of both application performance and security. Container technologies are used as the sandboxing mechanism for function executions by many commercial serverless platforms. Studies show that co-located function instances show effects of resource contention with regard to CPU and network bandwidth, raising concerns on the effectiveness of the isolation mechanisms \cite{wang2018peeking}. By design, components isolated by operating system (OS) level virtualization techniques (e.g.: containers), share hardware and the host’s OS kernel and thus are open to security vulnerabilities. AWS uses MicroVMs which are hardware-isolated lightweight virtual machines with their own mini-kernel \cite{Firecrac95:online}. They offer security and workload isolation from hardware-virtualization as with VMs, and the resource efficiency and smaller startup times of containers. Existing work in this area propose customized sandboxing techniques for function execution. Arguments are laid suggesting that stronger isolation mechanisms such as containers are needed only among different applications and weaker mechanisms such as having separate processes is sufficient for isolation among functions of the same application. Akkus et al. \cite{akkus2018sand} propose a new sandboxing method where different applications run inside separate containers, but different functions of the same application run inside the same container in parallel processes. Concurrent calls to the same function are handled inside the same container by spawning new processes by incrementing the memory allocation to the container. Since the memory footprint of a process is smaller than that of a container, this results in higher resource efficiency. Further, forking a new process inside a container only incurs a short startup latency. Oakes et al. \cite{oakes2018sock} present a container system optimized for serverless workloads. Bottlenecks of container cold starts are identified to be caused by limitations in linux isolation primitives and loading dependent packages.

\subsubsection{Billing Model}

Billing model refers to how costs are calculated for function executions in serverless environments. Since serverless offers a very fine-grained billing model, it is important to understand the fine details of billing criteria for different resources in order to make cost efficient resource allocation and scheduling decisions for both the provider and end user.

Serverless platforms offer a pay-as-you-execute billing model usually based on the CPU, memory and storage resource costs associated with each function execution. Additional costs may also be involved based on the nature of the application scenario and additional services consumed (e.g.:, State transitions costs of AWS step functions, Amazon Simple Queue Service (SQS) costs etc.). Although the basic billing model on the major commercial serverless platforms is the same, differences exist in finer details. Above its monthly free tier, AWS Lambda charges function executions per GB-s (rounded up to the nearest 1ms) of memory allocated and the number of execution requests \cite{AWSLambd95:online}. The billing scheme of Azure functions is similar to AWS except that billing is done per GB-s of average memory consumed during an execution instead of the memory allocated. Google functions charge users for both the GB-s of  memory provisioned and GHz-s of CPU provisioned.

Decisions on which functions of an application are suitable to be fused (combined) and which are best placed in the cloud or on the edge, are affected by the billing model \cite{elgamal2018costless}. Also, under a public/private hybrid cloud serverless model, function placement decisions in the private and the public cloud are based on the serverless billing model in the public cloud \cite{das2020skedulix}. Gunasekaran et al. \cite{gunasekaran2019spock} present a framework to use serverless functions and VM based executions interchangeably for better cost efficiency. They state that serverless functions are better suited during demand variations and lower average request rates while VMs are better with high request arrivals, due to the granular billing model in serverless platforms. Instead of the static pricing schemes of the existing platforms, schedulers could also utilize dynamic pricing schemes which enable users requiring higher QoS levels (e.g.: faster response time for applications with high delay sensitivity) to pay and acquire better services \cite{gupta2020utility}.

\subsubsection{Runtime Resource Limitations}

Serverless platforms impose various limitations on the allowed resource configurations for applications deployed on them. This is primarily aimed at improving flexibility in managing the available resources among multiple users without locking in a set of resources with a single user or application. These restrictions also prevent serverless systems in the public cloud from being subject to denial-of-service (DoS) attacks by malicious function requests flooding the resources attempting to overload the system and preventing legitimate requests from being fulfilled. On commercial serverless platforms, these limitations are primarily in the allowed memory, local disk space and cpu resource configurations, maximum allowed time for a function execution and the maximum number of parallel executions for a function without compromising on latency. AWS lambda allows users to select the amount of memory available to a function during execution from a range of 128-3008 MB, and allocates CPU power linearly in proportion to the configured memory. Concurrency is also allowed to be configured for each function with different options available which could be explored based on the workload \cite{AWSLambd95:online}. AWS lambda provides a non-persistent local disk of 512 MB \cite{wang2018peeking} and a maximum function timeout duration of 900 seconds. Azure functions on the other hand introduces different hosting plans for function apps such as the consumption, premium and dedicated plan, based on which the resource and time out limitations will be determined. Allowed maximum concurrency is 200 instances under all the plans \cite{AzureFun89:online}. For Google Cloud Functions, the maximum memory that a function can use is 4096 MB, while the function duration is capped at 540 seconds \cite{QuotasCl59:online}. Imposing limitations as discussed here, have detrimental effects for some application domains such as long running, compute intensive, data-driven applications \cite{hellerstein2018serverless}.

\subsection{Workload Management}

Since the serverless provider is responsible for the autonomic management of resources for applications, it is imperative that the platform develops an understanding of the nature, requirements and behavior of the incoming workloads. Three aspects in which awareness of the incoming workload is important for making efficient resource management decisions, are discussed below.

\subsubsection{Application Model}

Application model is the nature of the scheduling unit of an application that is deployed on a serverless platform. An application could have a certain structure and also QoS requirements that are independently specified for each task or for the application unit as a whole.

\par $\bullet$ \textit{Structure}: An application could be defined as a set of functions compiled in the form of a Directed Acyclic Graph (DAG), where each node is a function representing a fine-grained task and each edge represents a dependency among two functions \cite{carver2019search}, \cite{singhvi2019archipelago}, \cite{kim2020automated}, \cite{das2020skedulix}. A developer would specify the QoS requirements for a DAG based application either as a whole \cite{singhvi2019archipelago} or for each individual task. A serverless application could also be monolithic, composed of a single function, representing a single task \cite{hoseinyfarahabady2017qos}, \cite{aumala2019beyond}, \cite{stein2018adaptive}, \cite{mahmoudi2019optimizing}, \cite{mampage2021deadline}, \cite{kaffes2019centralized}. 

\par $\bullet$ \textit{Scheduling Granularity}: Scheduling granularity refers to the way in which scheduling algorithms handle the execution of an application submitted by a user. When an application takes the form of a DAG, it could either be scheduled as a single unit or each individual function could be scheduled separately. In addition, the scheduler could decide to queue requests and schedule them in batches \cite{zhang2020edge}. The scheduling granularity could be specified as a requirement of the developer or decided by a serverless platform, aiming for better resource efficiency.

AWS Lambda platform offers an orchestration mechanism for DAG based workflow executions, called AWS Step Functions. AWS treats user requests calling the first function in a workflow application and a request calling one function from another during a workflow execution, in a similar manner. Both requests go through the same scheduling policies \cite{AWSLambd95:online}. Azure Cloud Functions uses the concept of "function app", as the unit of scheduling and management of a serverless application. A function app is comprised of one or more functions with dependencies, which are scheduled, managed and scaled together. All the functions in a function app share the same pricing plan and runtime configurations  \cite{AzureFun89:online}. Studies suggest that orchestrating functions of the same application on an individual basis at a global system level could incur extra latency. Akkus et. al  \cite{akkus2018sand} suggest a serverless architecture utilizing a hierarchical message queuing mechanism with a local message bus on each host which handles local interactions among functions of the same application and their orchestration. Developers could also find the optimal way to fuse or combine a number of functions in a DAG for scheduling purposes, so as to reduce execution cost and latency associated with state transitions and network delays. This would be applicable to serverless deployments in the edge/fog environments as well due to enhanced delays over the network caused by data transfers across functions and associated costs \cite{elgamal2018costless}. When requests are less latency sensitive, the scheduler could decide to queue and schedule requests in batches or as a collection, for better resource efficiency.

\subsubsection{Workload Nature}

Serverless platforms are multi-tenant infrastructures and thus applications of multiple users compete for resources in a common shared environment. This could cause a lot of contention on the platform and lead to poor application performance if the resources are not properly managed as per the requirements of different applications. 

By nature any generic application could be either CPU, memory or I/O intensive. In some commercial serverless platforms, function placement on a VM is treated as a bin packing problem to maximize memory utilization \cite{wang2018peeking}. While this helps maximize resource utilization, it could also lead to CPU contentions if strict resource isolation strategies among applications are not adopted. Having an understanding on the rate at which requests arrive for applications too is an important factor.

The nature of the workload could be adopted in resource management decisions either when scheduling functions on VMs initially or in the subsequent management of limited resources. Mahmoudi et al. \cite{mahmoudi2019optimizing} use a machine learning based approach to develop a predictive performance model which tries to predict the normalized performance of any workload when assigned to a specific VM. Each time a new function is being deployed on a platform, a profiling step identifies the memory and CPU utilization levels of the function as well as the dependence on I/O features of the platform. Then predictive models are used to identify the VM that gives the best system performance for the function at that time slot. A good understanding of the nature of the applications is also important when the execution happens in an edge/cloud environment. Compute intensive functions are better served at the cloud which has seemingly unlimited resources even with a higher latency due to network delays. Lightweight functions may be better served at the edge with a lower response time. The placement decision of functions on the edge/cloud devices requires thorough understanding of the application behavior in each of the environments \cite{elgamal2018costless}, \cite{pinto2018dynamic}.

If a serverless platform  does not impose hard upper limits on the level of resources that a particular function instance could utilize, the co-location of many such functions having the same resource needs, could lead to deterioration of performance over time. Under such conditions, based on the state of the system and the resource needs of each application, dynamic management of applied resource limits to containers is an approach worth exploring more \cite{mampage2021deadline}, \cite{suresh2020ensure}. Further, we could use Linux Kernerl's cgroups to control and isolate resource usage of a collection of processes based on the requirements \cite{turner2010cpu}, \cite{kim2020automated}. These techniques could be adopted effectively with a better understanding on the needs of different workloads. Further, in addition to the resource requirements, modelling arrival rate of requests for different functions help in taking proactive resource scaling decisions \cite{singhvi2019archipelago}.

\subsubsection{Data Locality}

Data locality generally refers to the movement of computation to the nodes which contain the data required for the task execution. For data intensive applications this could improve the makespan drastically due to reduced network delays. Serverless platforms are known to decouple data management from function execution. This is a defining property of the serverless paradigm since the dis-aggregation of these two aspects allows serverless functions to scale in an independent manner \cite{garcia2019servermix}. For example, ensuring locality among serverless functions may mean executing functions which share data, in the same node or VM instance. While this would improve performance with fast shared memory, this could also reduce the flexibility of the provider to schedule functions and scale capacity \cite{garcia2019servermix}. Therefore scheduling and resource management techniques under the serverless model traditionally have not incorporated data-locality awareness in making their decisions. Applications are executed as stateless functions in commercial serverless platforms, and they share state through dis-aggregated storage services (e.g., Amazon S3) \cite{jonas2019cloud}, \cite{hellerstein2018serverless}. Thus serverless is said to be rather a data-shipping architecture (ships data to code) instead of shipping code to data \cite{hellerstein2018serverless}. As such, these delays in network and storage layers make this a less than ideal environment for latency and bandwidth sensitive, data-intensive applications such as machine learning. 

In existing works, many studies focus on load balancing methods to direct new function requests to workers with pro-actively spawned containers or containers that could be re-used \cite{singhvi2019archipelago}, \cite{ling2019pigeon}. Research works have also focused on attaining data locality among functions in terms of the runtime package dependencies of functions, by routing functions requiring similar packages to the same node for execution \cite{aumala2019beyond}. Lee et al. \cite{leegreedy} propose a greedy load balancing algorithm which tries to maximize locality and improve the cache-hit ratio while also minimizing load imbalance among nodes. The new sandboxing method proposed by Akkus et al. \cite{akkus2018sand} where functions of the same application share the same container, is able to improve latency by increased data locality, since the libraries shared by all the functions are needed to be loaded from memory only once.

\subsection{QoS Goal}

The resource allocation, scheduling and scaling techniques are aimed at satisfying certain requirements upon the execution of applications. These could be constraints that applications need to satisfy or an optimization goal that determines the performance of the resource management techniques.

\subsubsection{Latency}

Latency refers to the delay between a user request submission and the serverless provider's response. The request queuing time, resource set up time and the function execution time collectively result in the response time for a serverless application. The ability to maintain low latency for function executions is a key concern in serverless environments, specially since a majority of individual functions have execution times less than a second, or of a few seconds \cite{singhvi2019archipelago}. One main cause for high application latency in serverless environments is the cold start delay in resource setup, which tends to become significant compared to application execution times \cite{mampage2021deadline}. Scheduling algorithms along with different container pool management techniques and customized sandboxing methods to reduce resource setup times are explored in literature to reduce cold start delay \cite{aumala2019beyond}, \cite{singhvi2019archipelago}, \cite{ling2019pigeon}, \cite{gias2020cocoa}, \cite{akkus2018sand}, \cite{oakes2018sock}. Latency caused by CPU contention is also studied and dynamic resource control methods are proposed for serverless applications \cite{mampage2021deadline}, \cite{suresh2020ensure}, \cite{kim2020automated}, \cite{hoseinyfarahabady2017qos}.

\subsubsection{Throughput}

Throughput refers to the number of function requests processed by a serverless system in unit time. This is a good metric to demonstrate efficiency of the overall system and the ability to support high request arrival rates \cite{mahmoudi2019optimizing}. Throughput and mean latency of a system usually show a negative correlation. Thus techniques to reduce latency also improve throughput of a system. The allowed concurrency level for function instances in serverless platforms is also a determining factor for system throughput \cite{schuler2020ai}.

\subsubsection{User Cost}

Function execution cost to the user is as per the billing model of serverless platforms as described in the previous section. Since billing is mostly correlated with the function execution times, efforts to reduce execution latency result in optimized cost to the user. But for compute intensive applications, response time and cost could be inversely related since the allocated CPU time would affect the response time and also the execution cost. Experiments are done to observe the response time of functions for different resource allocations and associated costs. Such models could help a user decide on a suitable resource allocation scheme when there is a budget constraint and also the cost to achieve a certain performance level \cite{lin2020modeling}, \cite{gupta2020utility}. The scheduling granularity in terms of function fusion in serverless workflows could have an impact on the overall costs as well due to associated state transition costs and effects on function response times \cite{elgamal2018costless}.

\subsubsection{Resource Efficiency}

Resource efficiency refers to maintaining a high utilization level for the underlying active resources of the provider at all times. The fine-grained serverless billing model implies that the user is charged only for the resource-time actually consumed by the  application during its execution. Regardless of this, the provider has to maintain the overall infrastructure and as such, consolidating as many serverless applications as possible into a single host is in provider's best interest in order to yield better profits. On the other hand, packing too many requests on a single resource subsequently leads to poor performance. This reflects the typical conflicting objectives of the providers and consumers: minimization of cost and maximization of performance \cite{kim2020automated}. Thus finding an optimal resource consolidation strategy to the satisfaction of both parties is important. HoseinyFarahabady et al. \cite{hoseinyfarahabady2017qos} propose a QoS aware resource allocation controller for serverless environments which tries to minimize QoS violations to users while maintaining a healthy CPU utilization level of 60\%-80\% in each host in order to reach an ideal balance between system performance and energy consumption \cite{hoseinyfarahabady2017qos}. Resource efficiency is also important in serverless platforms deployed on edge resources as well due to limited capacities at the edge.

\subsubsection{Energy Efficiency}

Recently attention has also been given to the aspect of energy efficiency of running serverless systems, which is also connected to resource efficiency. Early surveys identify serverless computing to be promoting green computing due to the on-demand creation and release of resources used for function execution. Moreover, the model of billing per execution time incentivizes programmers to improve resource usage and execution time of their code \cite{shafiei2019serverless}. On the other hand, breaking down an application into a set of functions and the practice of setting up resources on-demand is deemed to cause additional latency and an execution overhead, which affect function performance. Kansi et al. \cite{kanso2017serverless} define a measure for energy efficiency, based on the mean time between two calls to the same function, mean time to setup resources and the mean time to execute the function. They show that the resource saving in serverless systems is generally highest when function calls are irregular and have large time gaps in between compared to the resource setup and function execution times. Deriving from this idea, Poth et al. \cite{poth2020sustainability} present a further developed model, capturing the overheads of the components which manage the serverless system during function invocation and execution, which is aimed at helping design decisions of serverless applications and systems. Gunasekaran et al. \cite{gunasekaran2020fifer} present a resource management framework to efficiently manage function chains on a serverless platform by improving container utilization and cluster wide energy consumption. 

\subsubsection{Security}

In addition to the security challenges that are common to any computing environment, serverless systems are susceptible to certain threats that are unique to these environments. Shafiei et al. \cite{shafiei2019serverless} identify application-level and function-level authentication to be one such specific challenge. Application-level
authentication refers to a mechanism which determines which users are allowed to access a certain application, while function-level authentication refers to the allowed invocations of one function from another. They also highlight the possible chance of replay attacks, where an intruder would capture a function execution request and execute it repeatedly, constraining the system resources and blocking legitimate users from accessing the service.  In general, runtime limitations are in place, in terms of limited execution times and maximum allowed CPU and memory allocations for a function, in order to minimize the effect of such attacks. Function isolation mechanism too plays an important role in enabling data privacy among tenants on the same host. Container namespaces are commonly used by current serverless systems for providing isolation among functions \cite{al2018making}.

\subsection{Elements of Resource Management}

All the factors discussed in detail in the above sections drive the development of techniques for the efficient management of resources in a serverless environment. In this section we briefly discuss the techniques explored in literature so far, under the three key elements of resource management that we have identified.

\subsubsection{Application workload Modelling}

Application workload modelling refers to using empirical and theoretical approaches to understand and model, the varying resource requirements of different applications,  workload arrival patterns and run time behavioral patterns. In the existing literature, mathematical modelling techniques are used in abundance in developing workload characterization techniques and prediction models in serverless environments. Singvi et al. \cite{singhvi2019archipelago} use a method of exponentially weighted moving average over the measured request arrival rate of the current interval and the previous estimate to get a new estimate of the workload arrival rates. Gunasekaran et al. \cite{gunasekaran2019spock} use a moving window Linear Regression (LR) model to predict the average request rate in their VM/FaaS hybrid model, in order to provision the required VMs in advance. Application execution latencies parameterized by the function input data sizes and framework overheads, are modeled using Regularized Ridge Regression in their work by Das et al. \cite{das2020skedulix}. They use training data obtained by running a substantial number of jobs in AWS Lambda and the OpenFaas platform deployed in the private cloud. Application specific performance models are developed in \cite{das2020performance} which are able to predict application latency and costs for various container configurations under a edge/cloud environment, considering network transfer times, container start up times, function execution times and storage latencies. They develop these using various techniques of regression over the training data sets. For estimating application latency in the cloud, they identify Gradient Boosted Regression Trees to be the most robust. These performance and cost models enable application developers to plan and budget their deployments in the most cost effective manner. Linear Regression is also used in \cite{gunasekaran2020fifer} for function execution time modelling which is then used in a Least Slack First (LSF) algorithm to select a request for execution from the queue. Zhang et al. \cite{zhang2020edge} use regression techniques to determine the total latency of executing a batch of serverless tasks over the edge and cloud runtime environments, in order to determine the runtime with the least latency. In their work, median sliding window time series modelling technique is used to predict runtime deployment time while Bayesian Ridge regression technique is used for processing time estimation. Attempts are also made in using ML based techniques such as k-means algorithm for clustering functions based on resource usage, for workload aware scheduling \cite{raith2021container}.

\subsubsection{Resource Scheduling}

Application scheduling in a serverless environment addresses the challenges of decision making with regard to resource provisioning, resource allocation and scheduling of function invocation requests. A comprehensive resource scheduling scheme would make use of performance prediction tools as identified above, for determining the optimal level of resource allocations and the scheduling policies to meet consumer and provider expectations. Constraint programming-based approaches try to minimize or maximize an objective by satisfying the constraints set for the function execution and the restrictions of available resources. Scheduling serverless functions over the underlying infrastructure of a serverless platform based on user SLA is a NP-hard problem. Therefore  algorithms and approaches for obtaining optimal scheduling decisions may not be feasible for these platforms considering the magnitude of the problem caused by the scale of resources. In contrast, using heuristic approaches is faster and they are scalable to large clusters. These approaches are able to provide acceptable performance and near-optimal solutions. Schedulers in commercial and open-source serverless platforms also seem to be using simple load balancing mechanisms such as round robin and bin-packing approaches \cite{wang2018peeking}, \cite{stein2018adaptive}. A considerable set of works exist in literature which study the applicability of various heuristic based approaches for function scheduling in serverless environments \cite{mampage2021deadline}, \cite{suresh2020ensure}, \cite{ling2019pigeon}, \cite{singhvi2019archipelago}. Although at very initial stages, learning based approaches such as machine learning, and deep reinforcement learning (DRL) methods are also being increasingly explored for resource allocation and scheduling decision making in these environments \cite{schuler2020ai}, \cite{mahmoudi2019optimizing}.

\subsubsection{Resource Scaling}

The ability to scale resources automatically to meet time varying application demand levels is a unique feature of paramount importance, under the serverless deployment model. In order to achieve high resource efficiency, the underlying resources are expected to scale up as required, when there is a rise of demand for an applications and scale down with diminishing demand. As such, an application would scale to zero with no resource consumption, at times when there is no demand for application execution. This ensures that the user is only billed for the exact resource consumption during application execution, relieving them of incurring costs for over provisioned idling resources during demand drops, which is often a critical issue in traditional serverful deployments. A good approach to scale resources also need to ensure that the QoS requirements of the user as well as the provider are sufficiently met.

Resource scaling or elasticity of resources in cloud, edge and fog environments, usually refers to two dimensions as horizontal and vertical scaling. Horizontal elasticity of resources allows to scale-up or scale-down the number of application instances, while vertical elasticity allows to adjust the amount of computing and other resources assigned to each application instance \cite{rossi2019horizontal}.

\par $\bullet$ \textit{Horizontal Scaling}: The existing commercial serverless platforms host containerized function instances in readily available VMs. Hence the horizontal scaling of serverless applications is affected by the availability of VMs with required resources. Experimentation done on AWS Lambda show that there is no significant change to cold start delay, when a new function instance is launched on a new VM previously not used for executions, and an existing VM \cite{wang2018peeking}. This indicates that the service providers usually maintain a pool of ready VMs for function executions and thus VM start up time is not usually a determining factor for function scheduling approaches. But recent interest in using the serverless model for compute intensive workloads such as predictive analysis applications using pre-trained deep learning models, present situations where dynamic scaling of VM resources is required to manage the dynamic workloads with varying resource requirements \cite{bhattacharjee2019barista}.

Serverless platforms often utilize containers as the sandboxing mechanism for the isolation of applications from each other. Each function instance is usually run on a separate container with the required resource configurations. Thus, prior to application execution, the required resource set up generally includes launching a new container, setting up the runtime environment and application specific initialization. The time taken for all these steps is collectively knows as the cold start latency. Although generally the coldstart latency of containers, which are lightweight resource units, is in the order of milliseconds, studies have shown that in serverless executions, this delay is largely dependent on each function's runtime dependencies and at times could grow to be even a few seconds \cite{wang2018peeking}, \cite{suresh2020ensure}. In order to attain the intended benefits of serverless auto-scaling abilities,  it is a necessity that the function cold start latencies are managed appropriately. 

To alleviate frequency of cold starts, serverless platforms often try to either reuse warm containers or create pre-warmed containers. The reuse of warm containers avoids any setup and initialization delays while pre-warmed containers avoid container launching delays. AWS, Azure and Google Cloud Functions maintain idle function instances for a particular time preiod, before they are recycled, in order to increase chances of container reuse \cite{wang2018peeking}. Apache OpenWhisk maintains pre-warmed containers and also uses load balancing strategies to direct similar function executions to the same set of workers as much as possible, thus enhancing chances of container reuse \cite{stein2018adaptive}. In existing research work, many models are presented to predict the arrival rates of incoming function requests and the demand for particular function executions and thereby proactively setup and maintain a pool of warm containers across VMs \cite{stein2018serverless}, \cite{singhvi2019archipelago}, \cite{ling2019pigeon}, \cite{bermbach2020using}, \cite{solaiman2020wlec}, \cite{xu2019adaptive}. Subsequently, load balancing algorithms are devised to benefit from these existing resource pools. In contrast to these works, Mohan et al. \cite{mohan2019agile} identify the processes of network creation and connection to be the major bottlenecks in container startup and propose a method for maintaining a pool of pre-created network elements which could be attached to a new function container whenever needed. This is done by using the concept of pause containers, which are network-ready empty containers which could be attached to other containers. Stein et al. \cite{stein2018serverless} propose a non-cooperative resource allocation heuristic for serverless environments which aims to predict the number of function instances required to be kept in order to maintain request waiting time below a threshold level. Gias et al. \cite{gias2020cocoa} compare the idle time of a function instance in a FaaS platform to the Time To Live (TTL) value of a TTL caching system and present a model to decide on the most suitable idle time that each function instances is to be maintained in the system so as to meet the function response time requirements of the user. At times the reuse of warm containers may even cause additional latencies for example when already fetched data in a function instance is out-of-date and required database connections have already reverted by the time a new request arrives at the container. Mechanisms to freshen up the warm instances prior to use is of importance in such instances \cite{hunhoff2020proactive}.

The concurrency level, which determines the maximum number of requests that the system could process in parallel for each different function, is also an important factor in function scaling. Existing commercial platforms have set fixed limits on concurrency levels for a particular function \cite{wang2018peeking}. Schuler et al. \cite{schuler2020ai} present a Q-learning based Deep Reinforcement Learning (DRL) approach to determine the best level of request concurrency, to achieve better performance in terms of system throughput and mean function latency.

\par $\bullet$ \textit{Vertical Scaling}: The core concept behind the serverless paradigm is to shift the complexities of application resource management from the developer to the service provider. Thus, the provider has the responsibility to autonomously manage application resource allocations as required, in contrast to allocating resources as per a detailed resource request under a serverful model \cite{jonas2019cloud}. For instance, AWS Lambda requires the user to only provide the amount of memory to be allocated per function instance and CPU power is stated to be allocated linearly in proportion to the requested memory \cite{AWSLambd95:online}. CPU is known to be a source of contention in serverless environments \cite{suresh2020ensure}, affecting application latencies. Hence, resource allocations to applications need to be monitored during  runtime, to avoid potential Service Level Agreement (SLA) violations to the user. Further, the providers need to carefully manage the CPU and memory resources allocated to applications in order to achieve resource efficiency and avoid over/under provisioning of resources. 

Adjusting CPU allocations to function instances in the runtime is a new research area in serverles computing. In the work of Suresh et al. \cite{suresh2020ensure}, they study the impact of dynamically adjusting CPU-shares to containerized function instances in the runtime. CPU-shares indicates the relative  weight given to a container in terms of the proportion of CPU  time it is given access to when CPU resources are limited \cite{Runtimeo76:online}. As per their model, containers are launched on VMs when spare memory capacity is available to accommodate the container. Thus multiple containers co-located on a VM would share the same processor core and thus the CPU time available to a container varies over time. This could hinder satisfying user latency requirements to certain applications. Experiments show that fine-tuned regulation of the CPU-shares allocations to containers dynamically could result in better QoS satisfaction. In a previous work \cite{mampage2021deadline}, we proposed a technique for dynamic CPU resource management to containers running serverless functions by applying the concept of cpu-quota  and  cpu-period  enabled  in  Linux Kernel’s  Completely  Fair  Scheduler  (CFS) \cite{turner2010cpu}. The cpu-quota value  sets the number of microseconds per cpu-period that the function instance's access to CPU resources is limited to, before it is throttled \cite{Runtimeo76:online}. Thus this acts as an effective ceiling and a hard limit for CPU resources allocated to a function instance. This technique helps in allocating a guaranteed CPU time for a function execution and also fine-grained management of underlying resources.

\section{Classification of Resource Management Techniques Using Taxonomy}

Based on the proposed taxonomy, we review existing key works on serverless resource management, as shown in Table \ref{tab:summary1}.

\begin{landscape}
\small

{\footnotesize
\begin{longtable}[c]{|m{2.2cm}|m{1.5cm}|m{1.3cm}|m{1.0cm}|m{1.0cm}|m{1.3cm}|m{1.3cm}|m{1.2cm}|m{1.4cm}|m{1.6cm}|m{1.3cm}|m{1.0cm}|}

\hline
 \multirow{3}{*}{\parbox{1.5cm}{\textbf{Work}}} &
  \multicolumn{3}{c|}{\textbf{System Design}} &
  \multicolumn{4}{c|}{\textbf{Workload   Management}} &
  \multirow{3}{*}{{\parbox{1.1cm}{\textbf{Quality of Service Goal}}}} &
  \multicolumn{3}{c|}{\multirow{2}{*}{\textbf{Resource Management Technique}}} \\ \cline{2-8}
  &

  \multirow{2}{*}{\begin{tabular}[c]{@{}l@{}}{\parbox{1.6cm} {\textbf{Deployment Model}}}\end{tabular}} &
  \multirow{2}{*}{\parbox{1.8cm}{\textbf{Application Isolation}}}  &
  \multirow{2}{*}{\begin{tabular}[c]{@{}l@{}}{\parbox{1.2cm}{\textbf{Pricing Model Awareness}}}\end{tabular}} &
  \multicolumn{2}{l|}{\textbf{Application   Model}} &
  \multirow{2}{*}{\begin{tabular}[c]{@{}l@{}}{\parbox{1.5cm}{\textbf{Workload   Nature Awareness}}}\end{tabular}} &
  \multirow{2}{*}{\parbox{1.3cm}{\textbf{Data \\Locality   Awareness}}}  &
  
   &
  \multicolumn{3}{c|}{} \\ 
  \cline{5-6} \cline{10-12}
   &
   &
   &
   &
  {\parbox{1.0cm}{\textbf{Structure}}}&
  {\begin{tabular}[c]{@{}l@{}}{\parbox{1.6cm}{\textbf{Scheduling Granularity}}}\end{tabular}} &
   &
   &
   &
   {\parbox{1cm}{\textbf{Application Workload Modelling}}}&
  {\parbox{1cm}{\textbf{Resource Scheduling}}} &
  {\parbox{1.5cm}{\textbf{Resource Scaling}}}\\
  \hline
\endhead

\cite{hoseinyfarahabady2017qos}  &
 \begin{tabular}[c]{@{}l@{}}Cloud\end{tabular} &
  Kernel thread &
  \hfil - &
  Task &
  Single Function &
  \hfil\checkmark &
  \hfil - &
  Latency,   Resource Efficiency &
  \hfil - &
  Feedback control system, PSO heuristic &
  \hfil - \\ \hline
\cite{stein2018serverless}  &
  Cloud &
  Container &
  \hfil - &
  Task &
  Single Function &
  \hfil\checkmark &
  \hfil - &
  Latency,   Resource Efficiency &
  \hfil - &
  Non-cooperative   game theoretic heuristic & 
  Heuristic \\ \hline
\cite{pinto2018dynamic} &
  Edge/Cloud &
  Container &
 \hfil - &
  Task &
  Single Function &
  \hfil\checkmark &
  \hfil - &
  Latency &
  \hfil - &
  Greedy,   UCB1, Bayesian UCB & 
  \hfil - \\ \hline
\cite{elgamal2018costless} &
  Edge/Cloud &
  Container &
  \hfil\checkmark &
  DAG &
  Function Fusion &
  \hfil\checkmark &
  \hfil\checkmark &
  Latency,   User Cost &
  \hfil - &
  LARAC &  
  \hfil - \\ \hline
\cite{mahmoudi2019optimizing}  &
  Cloud &
  Container &
  \hfil - &
  Task &
  Single Function &
  \hfil\checkmark &
  \hfil - &
  Latency,   Throughput &
  Machine   Learning &  
  Heuristic &
  Heuristic  \\ \hline
\cite{aumala2019beyond} &
  Cloud &
  Optimized container &
  \hfil - &
  Task &
  Single Function &
  \hfil - &
  \hfil\checkmark &
  Latency &
  \hfil - &
  Greedy & 
  \hfil - \\ \hline
\cite{singhvi2019archipelago}  &
  Cloud &
  Container &
  \hfil - &
  DAG &
  DAG, Single Function &
  \hfil\checkmark &
  \hfil\checkmark &
  Latency &
  Mathematical   Modelling &
  Heuristic &
  Heuristic \\ \hline
\cite{kaffes2019centralized} &
  Cloud &
  Container &
  \hfil - &
  Task &
  Single Function &
  \hfil - &
  \hfil - &
  Latency,   User Cost &
  \hfil - &
  Heuristic & 
  \hfil - \\ \hline
\cite{gunasekaran2019spock} &
  VM/Serverless based &
  Container &
  \hfil\checkmark &
  Task &
  Single Function &
  \hfil - &
  \hfil - &
  Latency,   User Cost &
  Mathematical   Modelling &
  Greedy &
  Heuristic \\ \hline
\cite{ling2019pigeon} &
  Cloud &
  Container &
  \hfil - &
  Task &
  Single Function &
  \hfil\checkmark &
  \hfil\checkmark &
  Latency,   Throughput &
  \hfil - &
  Greedy &  
  \hfil - \\ \hline
\cite{bhattacharjee2019barista} &
  Cloud &
  Container &
  \hfil\checkmark &
  Task &
  Single Function &
  \hfil\checkmark &
  \hfil\checkmark &
  Latency,   User Cost &
  Mathematical   Modelling,  Heuristic &
  Machine Learning &
  Machine Learning \\ \hline
\cite{suresh2020ensure} &
  Cloud &
  Container &
  \hfil - &
  Task &
  Single Function &
  \hfil\checkmark &
  \hfil - &
  Latency,   Resource Efficiency &
  \hfil - &
  Greedy &
  Heuristic  \\ \hline
\cite{das2020skedulix}  &
  Cloud &
  Container &
  \hfil\checkmark &
  DAG &
  Single Function &
  \hfil\checkmark &
  \hfil - &
  Latency,   User Cost &
  Mathematical   Modelling &
  Greedy &  
  \hfil - \\ \hline
\cite{kim2020automated} &
  Cloud &
  Process &
  \hfil - &
  Task/  DAG &
  Single Function &
  \hfil - &
  \hfil - &
  Latency,   Resource Efficiency &
  Feedback control system &
  Feedback control system, Heuristic &
  \hfil - \\ \hline
\cite{gupta2020utility}  &
  Cloud &
  \hfil - &
  \hfil\checkmark &
  DAG &
  Single Function &
  \hfil - &
  \hfil - &
  Latency &
  \hfil - &
  MILP &  
  \hfil - \\ \hline
\cite{tariq2020sequoia}   &
  Cloud &
  Container &
  \hfil - &
  Task/ DAG &
  DAG, Single Function &
  \hfil - &
  \hfil - &
  Throughput,   Resource Efficiency &
  \hfil - &
  Heuristics & 
  \hfil - \\ \hline
\cite{bermbach2020towards}  &
  Edge/Cloud &
  - &
  \hfil\checkmark &
  Task &
  Single Function &
  \hfil - &
  \hfil\checkmark &
  Resource   Efficiency &
  \hfil - &
  Heuristic &
  \hfil - \\ \hline
\cite{das2020performance} &
  Edge/Cloud &
  Container &
  \hfil\checkmark &
  Task &
  Single Function &
  \hfil\checkmark &
  \hfil - &
  Latency,   User Cost &
  Mathematical   Modelling &
  Heuristic & 
  \hfil - \\ \hline
\cite{gunasekaran2020fifer} &
  Cloud &
  Container &
  \hfil - &
  DAG &
  Single Function &
  \hfil\checkmark &
  \hfil\checkmark &
  Latency,   Resource   Efficiency,   Throughput &
  Mathematical   Modelling, ML   Modelling &
  Heuristics &
  Heuristic \\ \hline
\cite{rausch2021optimized} &
  Edge/Cloud &
  Container &
  \hfil\checkmark &
  Task &
  Single Function &
  \hfil\checkmark &
  \hfil\checkmark &
  Latency,   User Cost &
  \hfil - &
  Greedy & 
  \hfil - \\ \hline
\cite{zhang2020edge} &
  Edge/Cloud &
  Container &
  \hfil - &
  Task &
  Collection &
  \hfil\checkmark &
  \hfil - &
  Latency &
  Mathematical   Modelling &
  Heuristic &
  \hfil - \\ \hline
\cite{solaiman2020wlec} &
  Cloud &
  Container &
  \hfil - &
  Task &
  Single Function &
  \hfil - &
  \hfil\checkmark &
  Latency &
  \hfil - &
  Heuristic &
  Heuristic   \\ \hline
\cite{leegreedy} &
  Cloud &
  Container &
  \hfil -  &
  Task &
  Single Function &
  \hfil - &
  \hfil\checkmark &
  Latency &
  \hfil - &
  Greedy &
  Heuristic \\ \hline
\cite{mampage2021deadline} &
  Cloud &
  Container &
  \hfil -  &
  Task &
  Single Function &
  \hfil\checkmark &
  \hfil - &
  Resource Efficiency, Latency &
  \hfil - &
  Greedy &
  Heuristic \\ \hline

\addlinespace[2ex]
\caption{Classification of Resource Management Techniques } \label{tab:summary1}

\end{longtable}
}
\end{landscape}

\newpage

\section{Gap Analysis and Future Directions}

This detailed review on resource management under the emerging serverless computing model on edge, fog and cloud resources highlights open problems with great potential for exploration in future work. We discuss these areas in detail along the broader categories we have identified in this article, laying the groundwork for both research and development work in the coming years.

\subsection{System Design Characteristics}

Serverless model has been explored under different deployment models with hybrid infrastructures with geographical as well as performance distribution. These include a mix of edge, fog and cloud infrastructures along both private and public domains as well as hybrid deployment models realizing benefits from both serverless and traditional serverful models \cite{das2020skedulix}, \cite{gunasekaran2019spock}, \cite{pinto2018dynamic}. 

In the current commercial serverless platforms users are not allowed to specify the runtime environment for their function executions and also, they only offer access to CPU processing for the deployed applications. With increased exploitation of serverless for different application domains, there has been a rise in demand for access to specialized hardware where required. For example, deep learning models could benefit largely from GPU processing for inference. Research on such flexibile serverless frameworks is growing \cite{kim2018gpu}, \cite{naranjo2020accelerated} and has great potential.

One notable challenge for developers in adapting to the serverless model is the concern over the vendor lock-in effect due to the unique function signatures, naming conventions and other compatibilities required by each provider. Thus today, interest is building up on studies on multi-provider serverless support with reduced provider lock-in for applications \cite{aske2018supporting}.

The serverless model is built on the premise that the provider is at liberty to control infrastructure as they wish. This opens up interesting opportunities for them to seek maximum resource efficiency by leveraging idling resources from other running services as well as less attractive machines not conducive for leasing out to Infrastructure-as-a-Service (IaaS) customers \cite{jonas2019cloud}. Methods of consolidating and coordinating such resources for function deployments is a novel idea open for exploration.

To date the billing models in existing serverless platforms offer only fixed pricing schemes to all consumers. Customers having different requirements may be willing to pay more or less for enhanced or standard levels of service, calling for the need for dynamic pricing strategies \cite{gupta2020utility}.

\subsection{Workload Management and QoS Goals}

Serverless platforms still face many challenges when being explored for some applications with complex data flow dependencies.  Lack of an efficient function to function communication mechanism complicates the process of orchestrating complex workflow structures. Hierarchical message queuing and brokering mechanisms \cite{shafiei2019serverless} for handling local communication requirements among functions of the same application are interesting propositions but also have the downside of reducing extreme flexibility features offered by the whole serverless architecture. 

Further research could also be extended into providing more flexible QoS offerings to users, accompanied by suitable pricing models. Allowing too many specific demands from users could lead to poor resource efficiency for the provider and as such a compromise needs to be arrived at, after thorough investigation. Many resource scheduling models proposed for the serverless model focus on meeting the latency and budget constraints of consumers and rarely on the efficient resource usage at the provider. Maintenance of large resource pools for mitigating application latency deterioration would have very low attractiveness to a provider if it leads to a large resource wastage.

Modeling the aspect of energy efficiency in serverless environments in the cloud and more importantly in edge environments is a new avenue being explored from recent times, with a great potential for future work \cite{poth2020sustainability}.

\subsection{Resource Management Techniques}

Increased level of abstraction in specifying resource usage by developers leaves the serverless provider with the need to infer resource requirements and code dependency requirements in making appropriate resource allocation and scheduling decisions. Serverless platforms cater to needs of diverse applications with vastly differing characteristics and requirements. Understanding and profiling of workloads as per their resource needs and arrival patterns could largely help in taking scheduling decisions that could benefit both the consumers and providers. Research focus has already been directed towards modelling application performance and cost in serverless environments deployed in the cloud and at the edge, with different resource configurations for applications \cite{das2020performance}, \cite{lin2020modeling}. Although these models help in taking static scheduling decisions such as deciding on initial resource configurations and placement decisions, most of these approaches do not address dynamic factors that would affect performance in the runtime. Applications may not perform as required due to co-located application interference, machine performance variations and degradation due to over-utilization. Assessing intelligent methods for identifying and rectifying shortcomings in resource allocations in the runtime which aid in dynamic scaling decisions, is a research area with potential. 

Although initial efforts are made in extending serverless  model across the edge and fog computing networks, much further investigation is required in this area for taking effective resource management decisions. Given the involved delays across networks, transporting data and code to the edge for individual function executions as per the cloud serverless model would be ineffective specially for latency sensitive applications. Thus there is need for intermediary storage or caching services. Limitations casued by heterogeneity of edge and fog devices, on handling compute intensive function executions need to be considered in scheduling and scaling decisions. Data locality concerns inherent with the serverless model may be even more relevant for geographically distributed edge/fog computing networks.

Cold starts arising from the unique auto-scaling capability is a growing research area. Numerous solutions ranging from proactive resource scaling to reduce coldstart frequencies, to designing optimized sandbox solutions are proposed. A further interesting approach in this regard would be to incorporate intelligent techniques in deciding optimum function concurrency levels and methods for proactive resource provisioning to further diminish this challenge \cite{schuler2020ai}.

\section{Summary}

In this paper we have presented a comprehensive review on the aspect of resource management, referring to unique characteristics of this new serverless computing model. We propose a taxonomy covering the broader concept of resource management in edge, fog and cloud infrastructures, along the categories of system design decisions, approaches for incoming workload identification and management, and the QoS goals of involved parties. We also discuss three key aspects of resource management techniques and analyse the existing works using the proposed taxonomy. This taxonomy presents a clear view for serverless system designers on the essential features for consideration for a fully-fledged effective system. Further this provides a learning platform for researchers studying resource allocation, scheduling and scaling techniques in the serverless domain, based on which their work could demonstrate progress in the field. Finally, we provide a gap analysis referring to identified and addressed challenges, emphasizing on the vast potential for further research work.

%
\bibliographystyle{ACM-Reference-Format}
\bibliography{Serverless}

\end{document}